\documentclass[10pt,a4paper,twocolumn]{article}

\usepackage{color}
\usepackage{amsmath}
\usepackage{amssymb}
\usepackage{textcomp}
\usepackage{gensymb}
\usepackage{graphicx}
\usepackage{float}
\usepackage[utf8]{inputenc}
\usepackage{mathtools}
\usepackage{hyperref}

\title{Deep sub-arcsecond widefield imaging of the Lockman Hole field at $144\ \mathrm{MHz}$}
\author{\parbox{\linewidth}{\centering F. Sweijen$^{1\star}$, R. J. van Weeren$^{1}$, H. J. A. R\"ottgering$^{1}$, L. K. Morabito$^{2}$, N.Jackson$^{3}$, A. R. Offringa$^{4}$, S. van der Tol$^{4}$, B. Veenboer$^{4}$, J. B. R. Oonk$^{1,4,5}$, P. Best$^{6}$, M. Bondi$^{7}$,\\T. W. Shimwell$^{1,4}$, C. Tasse$^{8, 9, 10}$, A. P. Thomson$^{11}$}\\\\%
\parbox{\linewidth}{\centering%
{$\star$ e-mail: sweijen@strw.leidenuniv.nl}\\
\textit{\footnotesize$^{1}$Sterrewacht Leiden, University of Leiden, 2300 RA, Leiden, The Netherlands}\\
\textit{\footnotesize$^{2}$Centre for Extragalactic Astronomy, Department of Physics, Durham University, DH1 3LE, UK}\\
\textit{\footnotesize$^{3}$Jodrell Bank Centre for Astrophysics, School of Physics and Astronomy, University of Manchester, Manchester M13 9PL, UK}\\
\textit{\footnotesize$^{4}$ASTRON, Netherlands Institute for Radio Astronomy, Oude Hoogeveensedijk 4, Dwingeloo NL-7991 PD, The Netherlands}\\
\textit{\footnotesize$^5$SURF, P.O. Box 94613, 1090 GP Amsterdam, The Netherlands}\\
\textit{\footnotesize$^6$Institute for Astronomy, University of Edinburgh, Royal Observatory, Blackford Hill, Edinburgh, EH9 3HJ, UK}\\
\textit{\footnotesize$^{7}$INAF - Istituto di Radioastronomia, Via P. Gobetti 101, 40129, Bologna, Italy}\\
\textit{\footnotesize$^8$GEPI, Observatoire de Paris, CNRS, Université Paris Diderot, 5 place Jules Janssen, 92190 Meudon, France}\\
\textit{\footnotesize$^9$Centre for Radio Astronomy Techniques and Technologies, Department of Physics and Electronics, Rhodes University, Graham-stown 6140, South Africa}\\
\textit{\footnotesize$^{10}$USN, Observatoire de Paris, CNRS, PSL, UO, Nançay, France}\\
\textit{\footnotesize$^{11}$Jodrell Bank Centre for Astrophysics, The University of Manchester, SK11 9DL, UK}}}

\begin{document}
\maketitle
\textbf{High quality low-frequency radio surveys have the promise of advancing our understanding of many important topics in astrophysics, including the life cycle of active galactic nuclei (AGN), particle acceleration processes in jets, the history of star formation, and exoplanet magnetospheres. Currently leading low-frequency surveys reach an angular resolution of a few arcseconds. However, this resolution is not yet sufficient to study the more compact and distant sources in detail. Sub-arcsecond resolution is therefore the next milestone in advancing these fields. The biggest challenge at low radio frequencies is the ionosphere. If not adequately corrected for, ionospheric seeing blurs the images to arcsecond or even arcminute scales. Additionally, the required image size to map the degree-scale field of view of low-frequency radio telescopes at this resolution is far greater than what typical soft- and hardware is currently capable of handling. Here we present for the first time widefield sub-arcsecond imaging at low radio frequencies. We derive ionospheric corrections in a few dozen individual directions. Using a recently developed imaging algorithm \cite{Offringa2014,Tol2018,Veenboer2019,Veenboer2020} we then efficiently apply these corrections during imaging. This algorithm also reduces the computational cost of imaging, allowing us to efficiently map a large area of the sky. We demonstrate our method by applying it to an eight hour observation of the \textit{International LOw Frequency ARray} (LOFAR) \textit{Telescope} (ILT) \cite{Haarlem2013}. Doing so we have made a sensitive $7.4\ \mathrm{deg}^2$ $144\ \mathrm{MHz}$ map at a resolution of $0.3''$. The estimated $250,000$ core hours used to produce this image, fit comfortably in the budget of available computing facilities. This breakthrough will allow us to map the entire northern low-frequency sky at sub-arcsecond resolution.}

Endeavours in expanding sub-arcsecond resolution observations to both go deeper and wider have had successes at both 300 MHz \cite{Lenc2008} and GHz \cite{Garrett2001,Muxlow2005,Morgan2011,Chi2013,Middelberg2013,Radcliffe2019,Muxlow2020} frequencies using \textit{Very Long Baseline Interferometry} (VLBI). However, producing large contiguous images of significant sky areas (i.e. many degrees) at sub-arcsecond resolution, has not been done at low frequencies and is still tedious and time consuming at higher frequencies due to the limited field of view. This was further complicated due to a combination of a lack of known compact sources (limiting available calibrators and blind surveying), the long observing times required (limiting survey speed) or simply due to computational limits because of the amount of resources required to produce such images. Recent results, however, show that the vast majority of $150\ \mathrm{MHz}$ detected sources are still unresolved or barely resolved at the now routinely obtained angular resolution of $6''$ \cite{Shimwell2019}. The typical size of the $\mu$Jy population is of the order of a few kpc, which translates to sub-arcsecond angular sizes on the sky\cite{Bondi2018}. Sub-arcsecond resolution surveys will be able to start resolving these physical scales and are therefore an important and valuable resource.

Currently, an ideal instrument for this is the ILT. It is most sensitive at frequencies between $120$ and $168\ \mathrm{MHz}$. Stations throughout the European continent form baselines as short as $68\ \mathrm{m}$ and up to approximately $2000\ \mathrm{km}$. This results in a theoretical angular resolution of approximately $0.3''$ at $150\ \mathrm{MHz}$\footnote{The exact resolution varies depending on the visibility weighting scheme.}. The size of the international stations set the field of view at $2.6\degree$ full width at half maximum at $120\ \mathrm{MHz}$. Over the past decade innovative algorithms have been developed to overcome the ionospheric blurring sufficiently to allow high quality arcssecond $6''$ resolution imaging and novel computer science approaches have been designed to efficiently process data on large-scale parallel-compute infrastructure \cite{deGasperin2019}. Furthermore, recently developed pipelines have demonstrated our ability to produce low-frequency sub-arcsecond resolution images of individual sources that span regions of a few arcminutes \cite{Morabito2021}. Following this achievement, two major outstanding questions become: (i) can calibration of $\sim 10^3\ \mathrm{km}$ baselines be reliably extended to many sources, correcting for ionospheric effects at sub-arcsecond level across the full field of view and (ii) can we subsequently image the entire field of view at full resolution?

\begin{figure*}
    \centering
    \includegraphics[width=\textwidth]{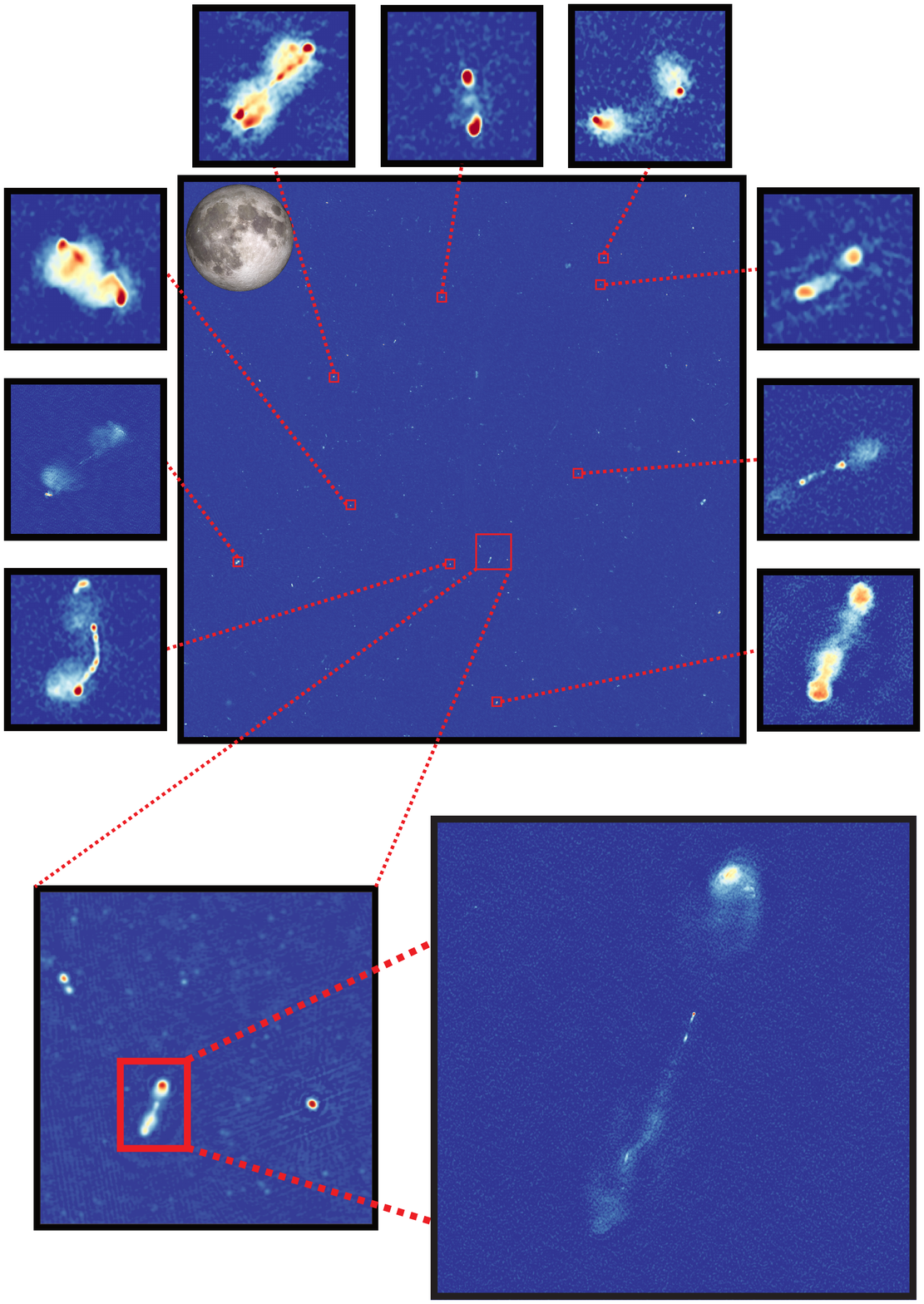}
    \caption{An overview of the field illustrating of both the widefield and high resolution aspects. The central image shows a $6''$ image of the covered part of field. The bottom two panels zoom in on a small portion of the field at $6''$ and a particular source at $0.3''$ in the bottom left and right, respectively. Surrounding the central image are highlights of various other sources detected in the $0.3''$ resolution map. Moon image credit: NASA.}
    \label{fig:mainimage}
\end{figure*}

Here we present our main result: the first deep widefield sub-arcsecond image at low radio frequencies. This image is shown in Fig.~\ref{fig:mainimage}. It covers $7.4\ \mathrm{deg}^2$ of the Lockman Hole field, centred at $\alpha=$10h45m00s, $\delta=$+58d05m00s, where there is a wealth of available ancillary data available (e.g. deep optical, near and far-infrared data). The ionospheric conditions during this observations were typical, based on the $6''$ image quality.

We demonstrate the feasibility of calibrating and imaging of the ILT's full international station field of view at sub-arcsecond resolution, providing a general strategy for widefield high-resolution imaging at low radio frequencies for dense interferometers such as the ILT.

The procedure we have developed to calibrate and image the full ILT field of view begins with a direction independent calibration on a known calibrator source (ILTJ104940+583529) from the Long Baseline Calibrator Survey (LBCS) \cite{Jackson2016,Jackson2021}. This direction-independent calibration is not sufficient to fully calibrate the FoV, however, and direction dependent effects (DDEs) remain. These mainly come from the ionosphere, which varies strongly over time and across the sky. Successfully correcting for these DDEs hinges on two conditions: the direction independent calibration needs to be approximately valid over the area that is to be calibrated and there need to be enough sources with a sufficiently high signal-to-noise ratio (SNR) in terms of compact structure to allow self-calibration to succeed. The former depends on the isoplanatic patch size while the latter depends on the amount of compact emission in sources.

Judging the amount of compact emission at sub-arcsecond scales in sources is difficult at a resolution of $6''$. Therefore, to facilitate the selection of DDE calibrator candidates, an intermediate resolution image with an angular resolution of approximately $1''$ was created. This gives a rough indication of the validity of the direction independent calibration across the field and of which sources are likely to retain compact emission at sub-arcsecond resolution. We selected sources that had a peak intensity of $25\ \mathrm{mJy}$ or higher in the $1''$ image, resulting in a sample of $46$ potential calibrators. The data were then phaseshifted to these sources, creating individual datasets, and self-calibrated.

In the direction-dependent self-calibration process we leverage the fact that the dominant remaining effects are the ionosphere and slowly varying effects, such as errors in the primary beam model. A starting model was automatically generated for each source by imaging it before the first iteration started. Subsequent self-calibration is an iterative process where in each iteration calibration solutions are obtained that provide an updated model, which can in turn provide better calibration solutions. Dominated by the ionosphere, we can constrain the ``phase-only'' self-calibration iterations to calibrate for differential Total Electron Content (dTEC), reducing the degrees of freedom by introducing the known frequency dependence of the ionospheric delay as the functional constraint given by Equation~\ref{eqn:tec}.
\begin{equation}
    \phi = -8.44797245 \times 10^9 \times \frac{\mathrm{dTEC}}{\nu},
    \label{eqn:tec}
\end{equation}
Here $\phi$ is the phase in radian and $\nu$ is the frequency in Hz. For every source, now  only a single dTEC value is determined per time interval, using the entire bandwidth, instead of frequency dependent phases. This reduces the degrees of freedom $48$ times compared to, for example, solving for scalar phases at a frequency interval of at $1\ \mathrm{MHz}$ (a typical solution interval for $6''$ calibration), significantly increasing the effective SNR. Once the ionosphere was calibrated for, long-timescale amplitude corrections were determined to solve remaining amplitude effects. For the ILT an imperfect primary beam model is one of the dominant causes of residual amplitude errors.

Manual inspection of the $46$ DDE calibrators concluded that $44$ converged after the dTEC iterations and $41$ allowed for subsequent amplitude corrections. The remaining $2$ sources did not have sufficient compact flux to perform self-calibration on and were thus discarded.

\begin{figure*}
    \centering
    \includegraphics[width=\textwidth]{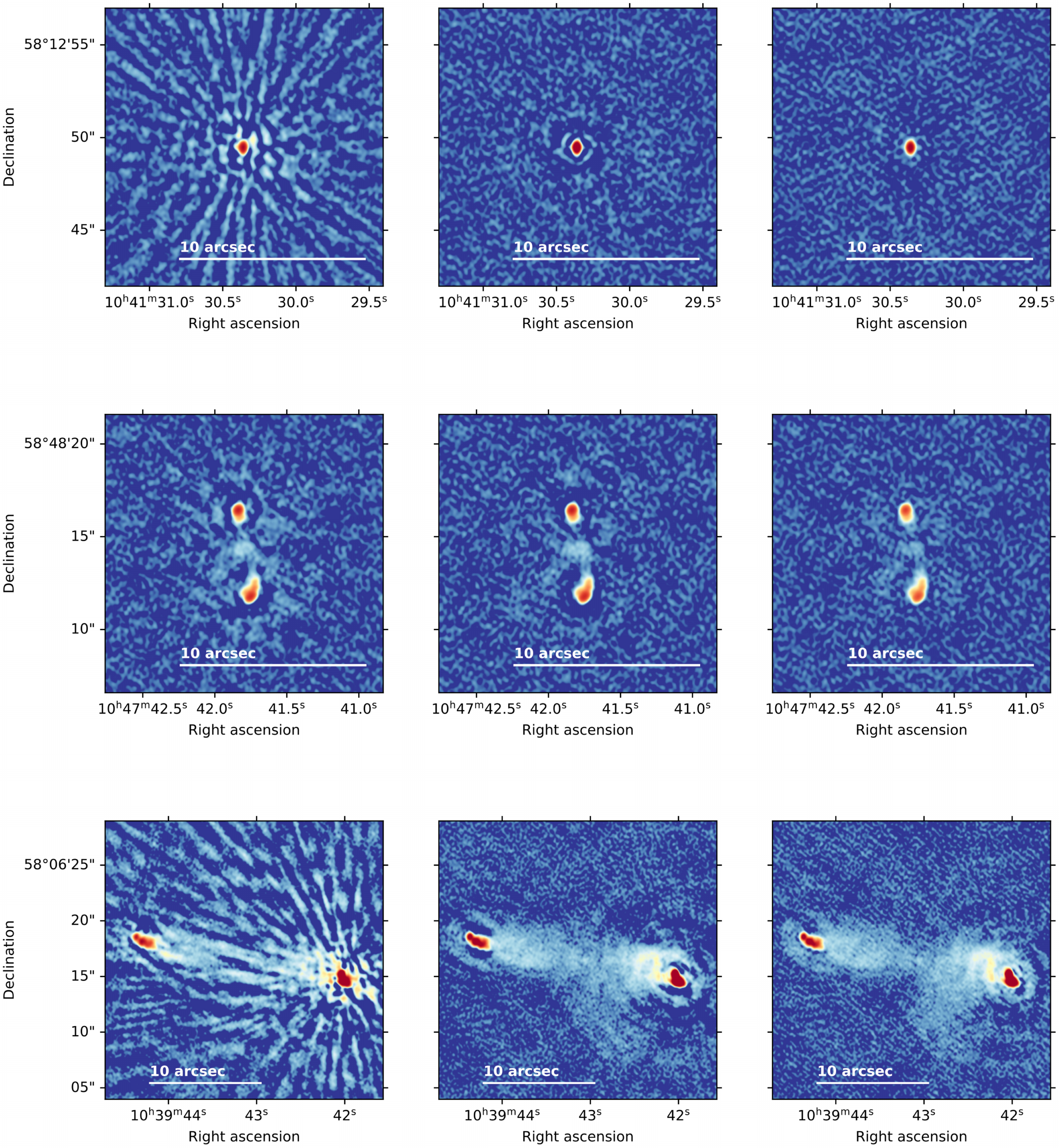}
    \caption{Illustration of the direction-dependent self-calibration process for different sources across the field and various degrees of extendedness. From left to right: the initial image with only direction-independent corrections applied, the image after dTEC corrections have been applied and finally the image after amplitude corrections have been applied. A significant improvement is seen after correcting ionospheric errors, while the amplitude corrections provide a more subtle improvement. The top and middle images cover a $15'' \times 15''$ area and the bottom image covers $25'' \times 25''$. A scale bar is shown for size.}
    \label{fig:ddeselfcal}
\end{figure*}

Figure~\ref{fig:ddeselfcal} shows the progression of self-calibration on three sources of varying complexity and distance from the initial LBCS calibrator. The fact that self-calibration converged for $44$ directions across the entire $7.4\ \mathrm{deg}^2$ area confirms that the density of compact sources is high enough to correct direction-dependent effects across the field of view.  Furthermore, the fact that a sufficiently accurate skymodel for initiating self-calibration could be created from a direction-independently calibrated dataset indicates that for similar ionospheric conditions a single correction is likely to be approximately valid over the entire field of view. Finally, the solutions were interpolated using radial basis function interpolation into a 2D map of corrections. This provides spatially smooth ``screens'', providing both dTEC and amplitude calibration solutions for every location in the field.

Even with perfect calibration, imaging the entire field of view is challenge of its own due to the sheer amount of pixels required. We chose a pixel scale of $0.11''\mathrm{px}^{-1}$ in order to Nyquist sample the main lobe of the PSF. Covering the ILT's field of view at $0.3''$ angular resolution would then require a total of $6-7 \times 10^9$ pixels. To put this in perspective, this is more than the entire \textit{Very Large Array} (VLA) \textit{Faint Images of the Radio Sky at Twenty-cm} (FIRST) survey which covers about $10^4$ square degrees using roughly $4 \times 10^{10}$ pixels \cite{White1997}.

At the time of writing it is not feasible to make a single image at the required size. Therefore the field was split in $25$ facets that were imaged individually. Each facet image covered a $0.69\degree \times 0.69\degree$ area that could be imaged at a manageable $22\,700 \times 22\,700$ pixels. For each facet, the $1''$ resolution image was used to subtract sources outside of the central $0.55\degree \times 0.55\degree$ region. This helps suppress artifacts that would arise from undeconvolved emission outside the image boundary. The imaged area extends beyond this subtraction border to account for remaining source structure due to imperfect subtraction. Lastly, the inner $0.55\degree \times 0.55\degree$ regions of the facets overlap with each other to avoid gaps in the mosaic. This process resulted in a final mosaic of approximately $83\ 950 \times 83\ 500$ pixels covering $7.4\ \mathrm{deg}^2$ on the sky.

Imaging was done using WSClean in combination with the newly developed Image Domain Gridder (IDG) \cite{Offringa2014,Tol2018,Veenboer2019}. Multi-frequency synthesis was used to account for the large bandwidth and multi-scale clean was used to account for both extended and compact emission. One of the bottlenecks for imaging is the need to place visibilities measured by a radio telescope on a regular grid in a computationally expensive process called gridding. The IDG algorithm is a novel approach to gridding that is computationally efficient as it circumvents the need to calculate expensive convolution functions during this process. It simultaneously allows for on-the-fly application of primary beam corrections and other direction-dependent corrections at little extra cost. Finally the IDG algorithm can leverage a GPU for the gridding calculations, giving a significant boost in speed compared to gridding on a CPU. Each of the facets took between $5$ and $7$ days of wall \textbf{clock} time to image. Using both local\footnote{ALICE; Leiden Observatory} and national\footnote{Spider; SURFsara} compute resources imaging ran in parallel and the effective wall time was significantly reduced.

A total of approximately $250\,000$ core hours were required for the data reduction process in order to arrive at the final image. This was split between calibration and imaging as $34\%$ and $76\%$, respectively. The image covers a $7.4\ \mathrm{deg}^2$ sky area at an angular resolution of $0.3'' \times 0.4''$ and a central rms noise level of $32\ \mu\mathrm{Jy\, beam}^{-1}$ is reached. This equates to a $1.4\ \mathrm{GHz}$ sensitivity of $7\ \mu\mathrm{Jy\, beam}^{-1}$, assuming a spectral index of $\alpha = -0.7$. This is comparable in depth to the deepest $1.4\ \mathrm{GHz}$ VLA image of the field \cite{Ivison2002}.

$7143$ sources are detected at $6''$ in the covered area using a single 8 hour observation with an rms noise level of $70\ \mathrm{uJy\ beam}^{-1}$. Of these $6397$ ($90\%$) are unresolved and $746$ ($10\%$) are resolved. Of those unresolved sources, $2568$ ($40\%$) had a high resolution counterpart detected at a SNR $> 5$. $293$ of these sources are resolved ($11\%$) and the other 2275 remain compact ($89\%$).

Our demonstration of wide-field, low-frequency, sub-arcsecond resolution imaging opens up important new scientific parameter space to be explored by the ability to conduct a blind all-sky sub-arcsecond survey. This resolution is a valuable tool to help, for example, disentangle radio-quiet AGN from star-forming galaxies, which exhibit similar unresolved radio properties and allow the study of radio sources nearby and far in new detail. Low-frequency data is especially important for broadband spectral modelling and at this resolution the ILT is well matched to other telescopes such as e-MERLIN or the VLA.

Over the past years the ILT has been conducting the LOFAR Two-Metre Sky Survey (LoTSS), a sensitive $6''$ survey of the northern sky \cite{Shimwell2017}. The data recorded for this survey holds incredible legacy value for a future sub-arcsecond survey. Over $75\%$ of the sky has already been observed for LoTSS, using the same duration, bandwidth and averaging as the data presented here. Weighted by integration time, only 3\% of this data contains no international stations. $54\%$ has $12$ or more international stations per hour of observing time and $84\%$ has $10$ or more international stations per hour of observing time. This means that a significant amount of data is already present in the LTA, ready to be processed. Relatively little extra or repeat observing will thus be necessary for a sub-arcsecond counterpart to LoTSS.

The work presented here demonstrates a general strategy for widefield low-frequency sub-arcsecond imaging. By combining advanced calibration strategies with new state of the art imaging techniques we have demonstrated complete direction-dependent calibration and widefield imaging at sub-arcsecond resolution at frequencies between $120$ and $168\ \mathrm{MHz}$. Our final result is a $7$ gigapixel image of the ILT's full international station field of view. This paves the road towards routine, pipelined data processing and a sub-arcsecond successor to the northern-sky LoTSS survey.

%TC:ignore
Correspondence and requests for materials should be addressed to F. Sweijen (sweijen@strw.leidenuniv.nl).

\clearpage

\clearpage
\section{Methods}
\subsection{Observation}
For this work we used an observation of the Lockman Hole field located at $\alpha = 10\mathrm{h}47\mathrm{m}00\mathrm{s}$, $\delta=58\degree05'00''$. This pointing was observed on July 12, 2018 from 11:08:10 to 19:08:10 UTC. In total 8 hours were spent on the target field, which was bookended by two 15 minute calibrator scans of 3C 196 and 3C 295, respectively. The observing setup was the standard LoTSS HBA setup with $48\ \mathrm{MHz}$ of bandwidth spanning a frequency range of $120$-$168$ MHz. With the exception of DE609, all $12$ other international stations partook in the observation. We will use the term \textit{Dutch LOFAR Telescope} (DLT) to refer to the core and remote stations.

\subsection{Calibration and imaging}
Calibration of the radio data started from the raw data stored in the Long Term Archive (LTA). First the data was processed using the de facto pipelines for calibrating the Dutch stations: \textsc{\textsc{prefactor}} and \textsc{\textsc{ddf-pipeline}}\cite{deGasperin2019,Shimwell2019,Tasse2021}. Subsequently a suitable infield \textit{Long Baseline Calibrator Survey} (LBCS) calibrator was selected. From thereon no dedicated pipeline existed. First direction independent calibration of the international stations was carried out. An intermediate resolution image at $\sim 1''$ was used to select sources for direction-dependent calibration, which were subsequently self-calibrated. Finally the field was split up in ``facets'' and imaged at the full native resolution.

\subsubsection{Direction independent calibration of the DLT}
The \textsc{prefactor} pipeline consists of two workflows for the calibrator scan and target scan respectively. A detailed overview of the calibration strategy is given by \cite{deGasperin2019}. A short overview of these workflows will be given next.

First the calibrator workflow determined the following corrections for both Dutch and international stations: a phase offset to align XX and YY polarisations (based on the assumption that the calibrator is unpolarised), a bandpass translating correlator units to physical units and a clock offset in nanoseconds to synchronize the clocks of each station to the same reference. These solutions were determined on a $4\ \mathrm{s}$ time interval and a $48.82\ \mathrm{kHz}$ frequency intverval. Finally they were transferred to the target dataset and the target workflow determined XX and YY phase solutions for the Dutch stations by calibrating against a skymodel from the TFIR GMRT Sky Survey (TGSS, \cite{Intema2017}) on an $8\ \mathrm{s}$ time interval and a $195.28\ \mathrm{kHz}$ frequency interval.

\subsubsection{Direction dependent calibration of the DLT}
Next the \textsc{ddf-pipeline} was run for direction-dependent calibration of the Dutch stations. See \cite{Shimwell2019,Tasse2018,Tasse2021} for a detailed description of this pipeline. First the direction-independent calibration was refined by means of self-calibration of the entire Dutch station field of view. Following this, both phase and amplitude calibration solutions were determined in $45$ directions across the Dutch field of view, following a facet-based approach. This provided a high-quality $6''$ resolution image of the target field covering the field of view of the Dutch stations.

\subsubsection{Direction independent calibration of the ILT}
To start calibration of the international stations, we followed the strategy as the \textsc{lofar-vlbi} pipeline \cite{Morabito2021}. Calibration solutions from \textsc{prefactor} and \textsc{ddf-pipeline} were applied, data was combined into $24$ manageable $2\ \mathrm{MHz}$ chunks of $32\ \mathrm{GB}$ and finally a suitable LBCS calibrator was selected within the international station field of view \cite{Jackson2021}. This calibrator source was then split off from the main dataset by phase shifting the visibilities towards this source. After phase shifting, the core stations were all combined into a single ``super station'', \texttt{ST001}. The dataset was also averaged down to a $16\ \mathrm{s}$ time interval and a $195.28\ \mathrm{kHz}$ frequency interval. This heavy averaging causes a $\sim 50\%$ intensity loss due to smearing at a distance of $\sim 4'$ from the phase centre, which considerably reduces the influence of other sources. ST001 has a field of view of the order of $\sim 4'$ as well, further significantly suppressing the effects of other sources on every baseline with this station. Self-calibration was performed using the routine described in \cite{Weeren2021}. This uses \textsc{DPPP} (\cite{Diepen2018}) to determine calibration solutions and WSClean (\cite{Offringa2014}) for imaging. First 9 ``fast'' iterations to solve for rapid phase variations were performed. The solution interval was $16\ \mathrm{s}$ (the resolution of the data) and the solution type was \texttt{scalarphase}, i.e. a phase as function of time and frequency, but independent of polarisation. Then 7 more iterations were performed where in addition to the fast solves, a ``slow'' solve using \textsc{DPPP}'s \texttt{complexgain} mode was done. This determined time, frequency and polarisation dependent amplitude and phase solutions on a $15\ \mathrm{min}$ timescale and a $195.28\ \mathrm{kHz}$ frequency interval, for the XX and YY polarisations. The starting model for the first phase-only iteration was a point source, due to the lack of more information, with the flux density taken from the low-resolution $6''$ image. While such a starting model is likely naive possibly incorrect, after each iteration the model was updated and the self-calibration cycle converged to believable double lobed source structure. These solutions were then applied to the target field dataset as initial direction-independent calibration for the international stations and refined direction-independent calibration solutions for the remote and core stations. The core stations were corrected with the solutions for \texttt{ST001}.

\subsubsection{Intermediate resolution imaging and faceting}
Due to its sheer size, the full field of view could not be covered with a single image at the ILT's native resolution. Therefore the field was split in $25$ facets that were imaged separately. To help suppress interference from undeconvolved sources outside of each facet and to determine suitable candidate sources for direction-dependent calibration, an intermediate resolution image at approximately $1''$ was made by tapering the data with a Gaussian taper. The resulting image was $25000 \times 25000$ pixels in size and covered the full field of view.

In preparation for high-resolution imaging, $25$ datasets ("facets") were then split out with their phase centres spread over the field of view in a $5 \times 5$ grid, with each facet spanning $0.55 \times 0.55\ \mathrm{deg}^2$ on the sky, separated by $0.5\ \mathrm{deg}$. For each facet the intermediate resolution map was subtracted outside this $0.55 \times 0.55\ \mathrm{deg}^2$ area. The $0.05\ \mathrm{deg}$ on each side ensured some overlap between the facets to not have gaps in the sky coverage. The data were then phase shifted to each of these phase centres and averaged to $4\ \mathrm{s}$ and $48.82\ \mathrm{kHz}$ time and frequency resolution. These were the final $25$ datasets that would be imaged at high resolution. The data for each facet were about $250\ \mathrm{GB}$ in total size.

\subsubsection{Direction dependent calibration of the ILT}
Sources that remain bright and compact at $1''$ resolution have a better chance of remaining compact down to $0.3''$ compared to a selection at $6''$. From the intermediate resolution image sources with a peak intensity above $25\ \mathrm{mJy/beam}$ were selected as candidate DDE calibrators. These were then split off into separate datasets similarly to the LBCS calibrator, but averaged more in time to a $1\ \mathrm{min}$ interval. For the brightest sources this balanced well the signal to noise ratio while still retaining the time resolution to correct most of the residual ionospheric distortions properly. The calibration solutions derived on the LBCS calibrator were pre-applied and a similar self-calibration routine as for the LBCS calibrator was subsequently used to self calibrate these sources, outlined below.
 
First an initial image was made from which a starting model was derived. As the majority of phase-related errors had been corrected on the infield calibrator, it was assumed that the remaining direction dependent effects were dominated by ionospheric differences. Therefore in the fast iterations \textsc{DPPP}'s \texttt{tec} mode was used to determine dTEC values rather than phases. In the slow iterations, \textsc{DPPP}'s \texttt{complexgain} mode was used to solve for remaining amplitude errors due to the slowly varying station beams. Solution intervals for both solves were calculated automatically based on the intensity detected on $800\ \mathrm{km}$ baselines \cite{Weeren2021}.
 
The \texttt{tec} iterations converged for $44$ sources and the \texttt{complexgain} iterations for $41$ sources. So-called ``jumps'', manifesting as solutions that are offset by a multiple of a specific ``jump'' value, were present in the \texttt{tec} solutions. These jumps are a consequence of local $\chi^2$ minima in the TEC-fitting \cite{Weeren2016}. This jump value was subtracted from the offending solution blocks iteratively until no jumps remained. The dTEC values were then spatially interpolated to a smooth screen over RA and DEC. A simple approach to convert discrete directions into screens would be a Voronoi tesselation based on the calibrator locations. This would, however, introduce sharp transitions, which would negatively impact IDG imaging later on. Therefore the solutions were interpolated with a multiquadric radial basis function instead, using SciPy's \texttt{interpolate.Rbf} function. Interpolation was constrained such that at the location of a calibrator source the screen solutions matched with the original solutions. Similarly the \texttt{complexgain} solutions were also interpolated to a screen. These two screens were used to apply direction-dependent calibration solutions during the final imaging.

\subsubsection{Final imaging}
Final high-resolution imaging of the data was done using WSClean in combination with IDG in CPU mode. Since the IDG algorithm can leverage a GPU for the gridding calculations, this amount can likely be significantly reduced\cite{Veenboer2020}. A $22500 \times 22500$ image of a similar dataset finished four times faster on a 22-core Intel (R) Xeon(R) Gold 6226 system with an Nvidia Tesla V100 compared to an identical system without one. Each of the facets was imaged separately with robust $-1$ weighting, a $0.11''$ pixel size and a $22700 \times 22700$ pixel image size. Multiscale multi-frequency synthesis clean was used together with spectral fitting using a second-order polynomial (through \texttt{-fit-spectral-pol 3}). Differential primary beam, dTEC and gain corrections were all applied on the fly during imaging. The beam corrections were applied every $10\ \mathrm{min}$, dTEC solutions were applied every minute and gain solutions every hour. This could be done efficiently through the use of IDG. The data was imaged using hardware on the Spider platform at surfSARA and the ALICE cluster in Leiden. Imaging took $5-7$ days wall clock time per facet on average using $24$ cores on a node with two 12-core Xeon Gold 6126 running at 2.6GHz and 384 GB of RAM. Effective wall clock time was significantly reduced however due to the availability of many compute nodes, allowing imaging to run in parallel  where each compute node imaged a facet independently. From start to finish the entire data reduction procedure consumed approximately $250,000$ core hours.

\subsection{Source detection and catalogues}
After imaging source detection was done using PyBDSF\cite{MohanRafferty2015}. Because source detection was done on each facet separately, there will be duplicate detections. Therefore the catalogues were first merged and then cleaned of duplicate detections. The compact sources and extended sources were treated separately for this. First we processed the compact sources, starting with a cross-match to the $6''$ resolution catalogue of the LOFAR Deep Fields survey with key optical host properties \cite{Kondapally2021,Tasse2021}. Next duplicate detections were removed using an internal cross-match within a radius of $0.15''$ of the fitted high-resolution position (i.e. approximately within the restoring beam of the high resolution map). The match closest to its facet centre was kept, while the others were discarded. Next a size constraint was introduced. Sources with fitted major or minor axes that were larger than $6''$ were discarded as poor fits as by definition the unresolved sources cannot be larger than the low resolution restoring beam. Next, a signal-to-noise ratio cut was made, constraining the peak intensity to be $\geq 5\sigma_\mathrm{rms}$, where $\sigma_\mathrm{rms}$ is the rms noise around the source as measured by PyBDSF. Finally, the compact and extended source catalogues were combined into a single catalogue and a final duplicate removal was done. The final catalogue contains $2483$ sources.

\subsection{Astrometric corrections}
As phases hold information about source positions, the self-calibration process may have shifted the astrometry away from a well defined reference. During calibration of the LBCS calibrator we noted self-calibration had ``recentered'' the image near one of the lobes, due to the use of a point-source starting model, introducing a noticeable offset that needs correcting. Therefore the final image needs to be realigned to a reference. As our reference we used the final calibrated and aligned radio catalogue of Lockman Hole from the LOFAR Deep Fields survey.

Offsets to align the high resolution image to the low resolution image were determined for each facet separately. First, point sources with $\mathrm{SNR} \coloneqq I_\mathrm{peak} / \sigma_\mathrm{rms}^\mathrm{local} > 5$ were selected. Secondly an isolation constraint for no other sources to be present within $30''$ was applied. Astrometric offsets were then determined by the median value of the difference with the $6''$ position for right ascension and declination, respectively. Supplementary Figure 1 shows the offsets for each facet. There is a roughly systematic offset due to the aforementioned recentring behaviour during calibration of the LBCS source. The remaining differences are due to additional offsets in those facets, but are small and of the order of one resolution element or less. 

A single offset using only $\mathrm{SNR} > 25$ sources was also determined as a check. The median values obtained from this agree with the median of all individual facets combined. Therefore the facet-based offsets are adopted to allow for inter-facet variations.

\subsection{Smearing}
A variety of effects can degrade the effective resolution in the image, referred to as the ``smearing'' of sources. This results in a reduction of peak intensity and artificial broadening of sources. These effects include time and bandwidth smearing, ionospheric disturbances and interstellar or interplanetary scattering (ISS/IPS). Time and bandwidth smearing is an ``instrumental'' effect that occurs when averaging the data. It artificially broadens sources in the image radially (bandwidth smearing) and azimuthally (time smearing). Supplementary Figure 2 illustrates the smearing losses on the longest baseline\cite{Bridle1999}. Ionospheric disturbances are mostly corrected for in both the direction-independent and direction-dependent calibration process, but are difficult to perfectly remove as they are limited by the density of DD calibrators, available SNR, solution intervals and source complexity. The result is similar to atmospheric seeing. Finally interstellar and interplanetary scattering scramble astronomical radio signals before they even reach the ionosphere and hence imprint an inherent, uncorrectable smearing on the data\cite{Rickett1990}. While the former can technically be resolved with better calibration, the latter sets a fundamental limit to the effective resolution that can be achieved. The combination of ionospheric seeing and interstellar and -planetary scattering manifests itself as the broadening of sources near the phase centre, where time and bandwidth smearing is non-existent.

If smearing is not accounted for, peak intensity and intrinsic size measurements will be off due to a wrong assumed PSF. Measuring the most compact sources in the centre of the field, where time and bandwidth smearing is negligible, will indicate whether broadening from ISS/IPS is an issue for this observation. The estimated PSF from WSClean is $0.38'' \times 0.30''$, while the smallest measured source in the central facet is $0.39'' \times 0.31''$. The expected broadening from ISS/IPS is $\sim 50\ \mathrm{masec}$\cite{Cohen1974}. We thus conclude that we achieve the theoretical angular resolution with possible minor broadening from ISS/IPS.

Broadening from time and bandwidth smearing will vary across the field. The net effect is a combination of the local PSF and smearing. We give the theoretical reduction in peak intensity as a measure of how severe the smearing is in Supplementary Figure 2. For accurate deconvolved source sizes the smearing would need to be taken into account. A significant reduction in time smearing can be achieved by processing the data at its archived $1\ \mathrm{s}$ time resolution.

\subsection{Flux scale corrections}
A correction to the flux scale was necessary, as the reference bandpasses that were used were not tied to any particular flux density scale. For a reliable flux scale correction sources with the same flux density at low and high resolution were needed. A selection was made based on high-significance sources in the high-resolution image. Compact single-gaussian sources with $\mathrm{SNR} > 25$ were selected and the median ratio $S_\mathrm{ILT} / S_\mathrm{DLT}$ was determined. The selection of bright high-SNR and compact sources ensures no flux density is lost. Smearing effects are mitigated by using the flux density, which should be conserved, instead of peak intensity. The correction factor was determined to be $1.21 \pm 0.19$ from the median flux density ratio. After scaling, the mean ratio is $S_\mathrm{ILT} / S_\mathrm{DLT} = 0.99_{-0.34}^{+0.26}$ and the median ratio is $S_\mathrm{ILT} / S_\mathrm{DLT} = 1.00_{-0.18}^{+0.14}\ (0.82$ -- $1.14)$. Both the uncertainty and the values in parentheses reflect the $16^\mathrm{th}$ and $84^\mathrm{th}$ percentiles, respectively. The distribution of the ratios over all facets after scaling can be seen in Fig.~\ref{fig:astrometry}. A small number of sources is seen to have a ratio significantly smaller than unity. These are most likely sources that have extended emission below the detection threshold. This is a small number of outlying sources however, and has not affected the scaling in a significant way.

The uncertainty on measured flux densities is compounded by two main effects: an uncertainty in the general flux density scale ($\sigma_\mathrm{cal}$) and an uncertainty in our scaling factor ($\sigma_\mathrm{scale}$). For the former we assume a conservative error of $20\%$\cite{Shimwell2019}. This was added in quadrature with the other uncertainty to arrive at the total uncertainty on the flux density of $\sigma_\mathrm{total} = \sqrt{\sigma_\mathrm{cal}^2 + \sigma_\mathrm{scale}^2} = 30\%$.

\section*{Acknowledgements}
We thank the (anonymous) reviewers for their comments. FS would like to thank SURFsara for beta access to the Spider platform. This paper is based (in part) on data obtained with the International LOFAR Telescope (ILT) under project code LT10\_012. LOFAR (van Haarlem et al. 2013) is the Low Frequency Array designed and constructed by ASTRON. It has observing, data processing, and data storage facilities in several countries, that are owned by various parties (each with their own funding sources), and that are collectively operated by the ILT foundation under a joint scientific policy. The ILT resources have benefitted from the following recent major funding sources: CNRS-INSU, Observatoire de Paris and Université d'Orléans, France; BMBF, MIWF-NRW, MPG, Germany; Science Foundation Ireland (SFI), Department of Business, Enterprise and Innovation (DBEI), Ireland; NWO, The Netherlands; The Science and Technology Facilities Council, UK[7]. RJvW acknowledges support from the ERC Starting Grant ClusterWeb 804208. This work was supported by the Medical Research Council [grant MR/T042842/1]. This work made use of the Dutch national e-infrastructure with the support of the SURF Cooperative using grant no. EINF-251. PNB is grateful for support from the UK STFC via grants ST/R000972/1 and ST/V000594/1. This work was performed using the compute resources from the Academic Leiden Interdisciplinary Cluster Environment (ALICE) provided by Leiden University. This work also made use of SciPy, NumPy, Matplotlib and TOPCAT.

\section*{Author contribution statements}
FS led the paper, reduced the data and produced the images. RJvW developed the self-calibration routine and helped expanding it to international baselines. HJAR helped scope and write the paper. LKM and NJ conducted the LBCS survey and developed the \textsc{lofar-vlbi} pipeline which served as the foundation for this work. ARO, SvdT and BV maintain WSClean, IDG and were of great help in fixing problems and adding features to the software. JBRO helped secure resources on SURFsara and provide support for our data reduction on their Spider platform. PNB led the proposal that obtained the data and contributed to editing of the paper. MB is a member of the long-baseline working group and contributed to editing of the paper. TWS contributed towards some of the various data processing pipelines used in this work and produced the deep Lockman Hole 6'' image. CT developed the DDFacet software used and produced the deep Lockman Hole 6'' image. APT is a member of the long-baseline working group and helped determining the flux scale.

\section*{Competing interest statement}
The authors declare no competing interests.

\section*{Data Availability}
A source catalogue and the full-resolution images are accessible through the LOFAR Surveys Key Science Project webpage: \url{https://www.lofar-surveys.org/hdfields.html}

\section*{Code Availability}
The various software and pipelines used in this work are publicly available at \url{https://github.com/lofar-astron/prefactor} (\textsc{prefactor}), \url{https://github.com/cyriltasse/DDFacet} (\textsc{ddfacet}), \url{https://github.com/lmorabit/lofar-vlbi} (\textsc{lofar-vlbi}), \url{https://github.com/mhardcastle/ddf-pipeline} (\textsc{ddf-pipeline}) and \url{https://gitlab.com/astron-idg/idg} (IDG). Code from other parts not using these pipelines is not directly available, because it is not part of a complete pipeline, but is available upon reasonable request to the authors.

\clearpage
\setcounter{figure}{0}
\renewcommand{\figurename}{Supplementary Figure}
\begin{figure*}
    \centering
    \includegraphics[width=\textwidth]{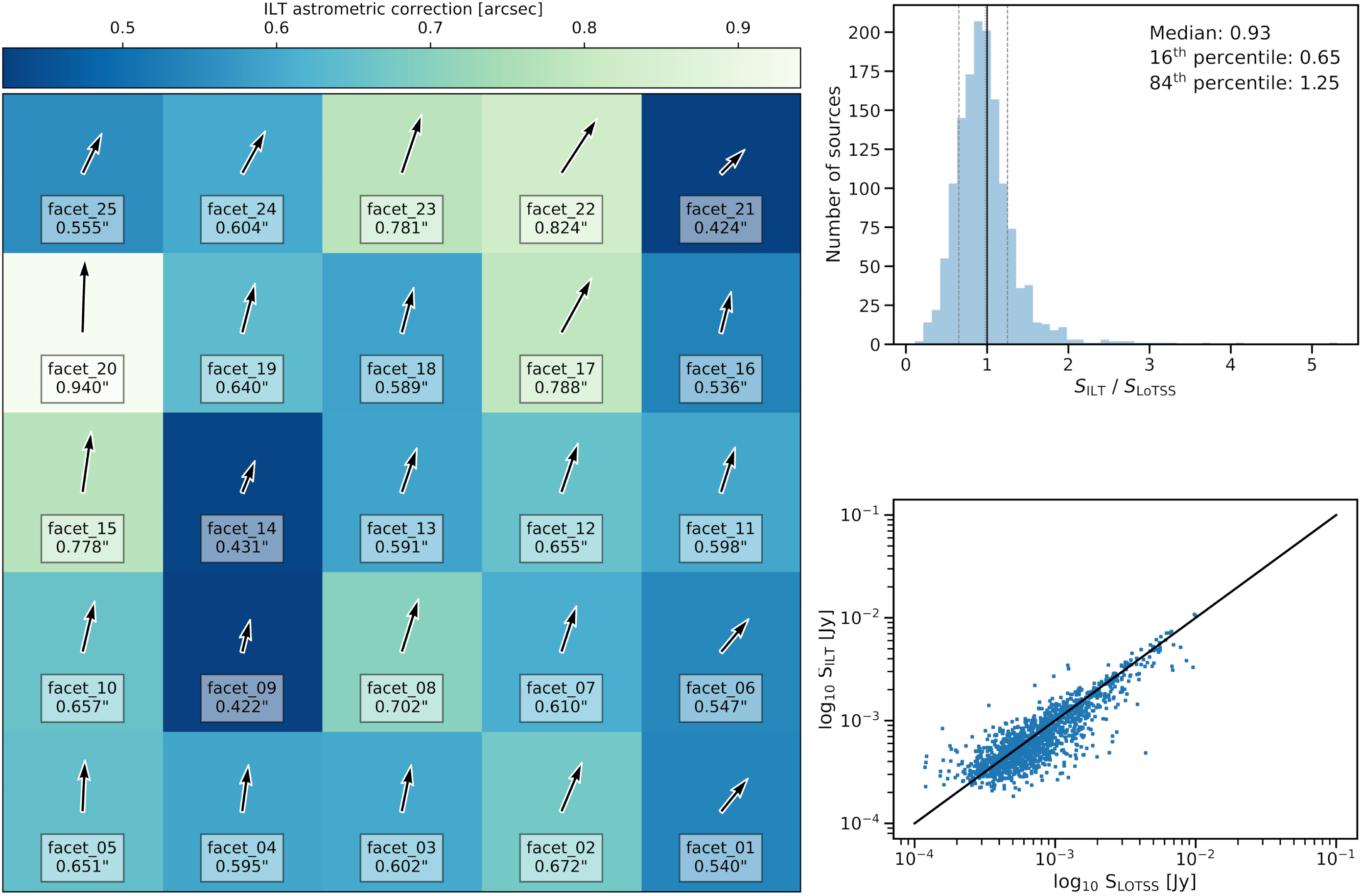}
    \caption{Astrometric offsets per facet and flux density distributions after setting the absolute flux scale. \textit{Left:} Offsets in arcsecond to align the high-resolution catalogue with the $6''$ catalogue. Each square is one facet. The black arrows are to scale with respect to each other and indicate the magnitude and direction of the applied offset. \textit{Top right}: The distribution of flux density ratios of compact SNR $> 5$ sources, after scaling. The median, $16^\mathrm{th}$ and $86^\mathrm{th}$ percentiles are indicated by the dotted and dashed gray lines, respectively. The black line indicates unity. \textit{Bottom right}: Flux densities measured with the international array versus those measured with the Dutch array. The black line indicates unity.}
    \label{fig:astrometry}
\end{figure*}

\begin{figure*}
    \centering
    \includegraphics[width=\textwidth]{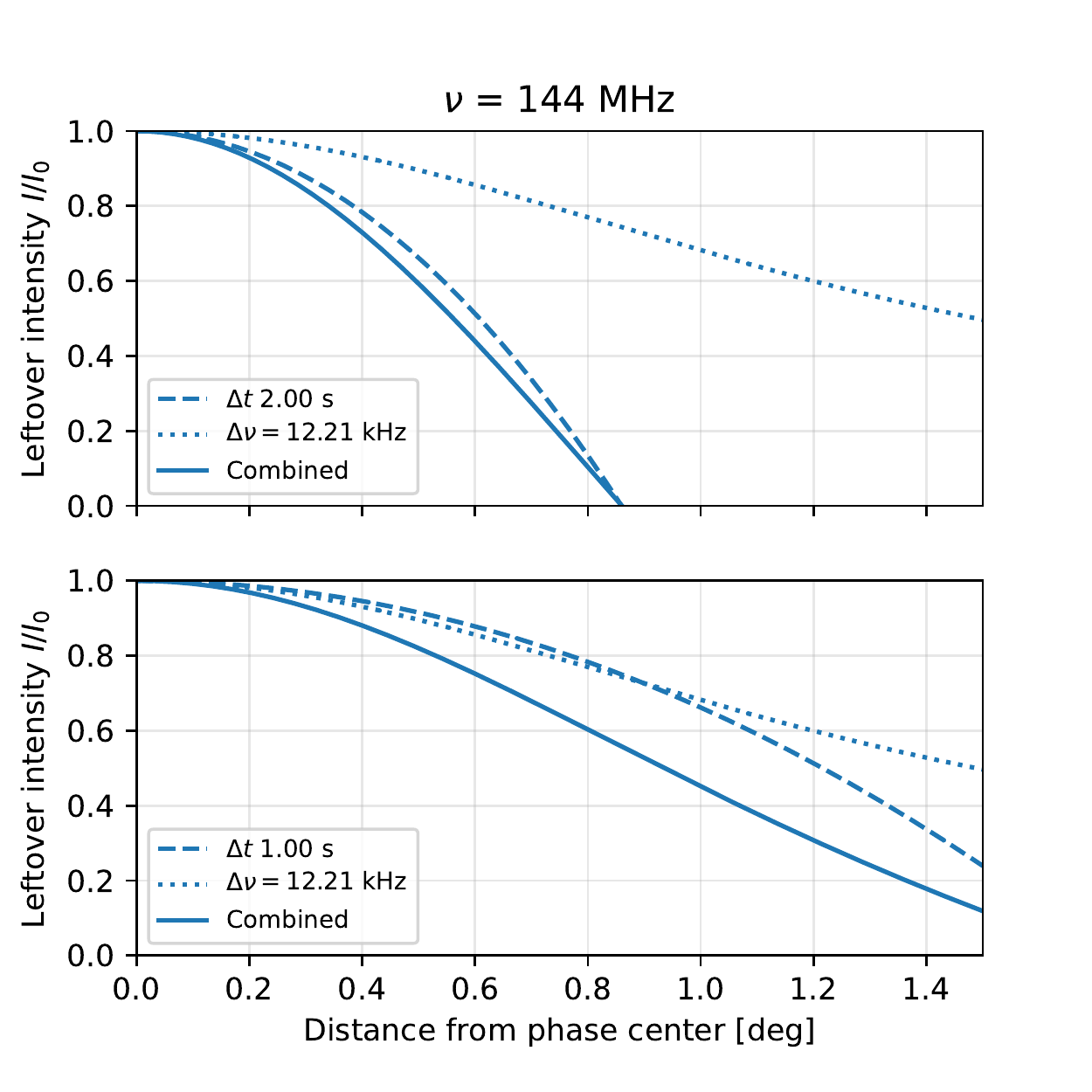}\\
    \caption{Time and bandwidth smearing losses on the longest baseline ($1986\ \mathrm{km}$, $\theta = 0.2''$ at $144\ \mathrm{MHz}$), for the central frequency, as a function of distance to the phase centre. Losses are expressed as the remaining peak intensity $I$ as a fraction of the original peak intensity $I_0$. The dotted line indicates the reduction in peak intensity from bandwidth smearing, the dashed line the reduction from time smearing and the solid line the total reduction from both effects. The top panel shows smearing at the averaging of this dataset, while the bottom panel shows the smearing at the archived averaging parameters.}
    \label{fig:smearingandflux}
\end{figure*}
%TC:endignore
\clearpage
\bibliographystyle{plain}
\bibliography{references}
\end{document}